\documentclass[11pt]{article}
\usepackage[english]{babel}

\usepackage[a4paper,top=2cm,bottom=2cm,left=2.5cm,right=2.5cm,marginparwidth=1.75cm]{geometry}
\usepackage{setspace}
\usepackage{amsmath}
\usepackage{graphicx}
\usepackage[colorlinks=true, allcolors=blue]{hyperref}

\title{GALATEA: The 15-m Galactic Archaeology Spectroscopic Surveyor}

\author{
  Borja Anguiano$^{1}$, First Author$^{2}$, Second Author$^{3}$\\[0.5em]
  {\small
  $^{1}$Centro de Estudios de F\'isica del Cosmos de Arag\'on (CEFCA), Spain\\
  $^{2}$Affiliation 2 (ESO Member State)\\
  $^{3}$Affiliation 3 (ESO Member State)
  }
}

\date{\today}

\usepackage{etoolbox}   

\makeatletter
\patchcmd{\thebibliography}
  {\list}
  {\small%
   \list}
  {}{}

\patchcmd{\thebibliography}
  {\setlength{\itemsep}{\z@}}
  {\setlength{\itemsep}{0pt}%
   \setlength{\parsep}{0pt}}
  {}{}
\makeatother

\begin{document}

\begin{titlepage}
    \centering

    {\Large ESO Expanding Horizons Call \par}
    \vspace{0.8cm}

    {\LARGE\bfseries GALATEA: The 15-m Galactic Archaeology Spectroscopic Surveyor\par}
    \vspace{1.5cm}

    {\large
    Borja Anguiano$^{1}$, David Valls-Gabaud$^{2}$, Andr\'es del Pino$^{3}$, Guillaume F. Thomas$^{4,5}$, Alberto M. Mart\'inez-Garc\'ia$^{3}$, Ivan Minchev$^{6}$, Patricia Sanchez-Blazquez$^{7}$, Danny Horta$^{8}$\par}
    \vspace{0.5cm}

    {\small
    $^{1}$Centro de Estudios de F\'isica del Cosmos de Arag\'on (CEFCA), Plaza San Juan 1, 44001 Teruel, Spain.\\
    $^{2}$LERMA, CNRS, Observatoire de Paris, 61 Avenue de l'Observatoire, 75014 Paris, France.\\
    $^{3}$Instituto de Astrof\'isica de Andaluc\'ia (IAA-CSIC), Glorieta de la Astronom\'ia, 18080 Granada, Spain.\\
    $^{4}$Instituto de Astrof\'isica de Canarias (IAC), 38205 La Laguna, Tenerife, Spain.\\
    $^{5}$Universidad de La Laguna, Dpto. Astrof\'isica, 38206 La Laguna, Tenerife, Spain.\\ 
    $^{6}$Leibniz-Institut f\"ur Astrophysik Potsdam (AIP), An der Sternwarte 16, 14482 Potsdam, Germany.\\
    $^{7}$Departamento de F\'isica de la Tierra y Astrof\'isica \& IPARCOS, Universidad Complutense de Madrid, 28040 Madrid, Spain.\\
    $^{8}$Institute for Astronomy, University of Edinburgh, Royal Observatory, Blackford Hill, Edinburgh EH9 3HJ, UK\par}
    \vspace{0.8cm}

    {\small
    E-mail: \texttt{banguiano@cefca.es}
    \par}

    \vfill

    {\small White paper submitted to the ESO ``Expanding Horizons'' Call\par}
    \vspace{0.5cm}

    {\large \today\par}
\end{titlepage}

\setcounter{page}{1}


\begin{abstract}
\textsc{GALATEA} (the \emph{Galactic Archaeology and Local-group Astrophysics Telescope for Extended Areas}) is a concept for a dedicated 15-m, wide-field, 10,000-fibre spectroscopic survey facility in the northern hemisphere, optimized for degree-scale, multi-object spectroscopy. With a $\sim 1~\mathrm{deg}^2$ corrected field-of-view and both medium- ($R \simeq 5{,}000$--$10{,}000$) and high-resolution ($R \simeq 20{,}000$--$25{,}000$) modes, \textsc{GALATEA} would open a new regime in Galactic and Local Group astronomy: deep, chemically detailed spectroscopy of vast samples of individual stars in the outer disc, warp, flare, halo substructures, M31, M33 and their dwarf satellites, far beyond the reach of current surveys. By delivering precise radial velocities and detailed chemical abundances for stars with exquisite astrometry and photometry from \emph{Gaia} and its proposed near-infrared successor \emph{GaiaNIR}, \textsc{GALATEA} will complete and fully exploit the 6D phase-space and chemodynamical information for these populations. Compared to existing northern multi-object spectroscopic facilities (BOSS, APOGEE, DESI, LAMOST, WEAVE, PFS), \textsc{GALATEA} delivers an order-of-magnitude jump in survey power ($\propto D^{2} \times N_{\mathrm{fibres}}$) by combining a 15-m aperture, $\sim 1~\mathrm{deg}^2$ field, and 10{,}000 fibres in a single dedicated facility. It is also strongly complementary to 30--40\,m ELTs: GALATEA provides the wide-field, high-multiplex discovery and chemodynamical mapping, while ELTs deliver deep, high-resolution follow-up of the faintest or most complex targets.

\end{abstract}

\section{Introduction and Motivation}

The coming decade will be defined by an unprecedented wealth of astrometric, photometric, time-domain and multi-messenger data from facilities such as \emph{Gaia}~[1], the proposed near-IR mission \emph{GaiaNIR}~[2], LSST~[3], \emph{Euclid}~[4], \emph{PLATO}~[5], \emph{LISA}~[6], \emph{NewAthena}~[7], the \emph{Nancy Grace Roman Space Telescope}~[8]. Turning these surveys into a quantitative chemodynamical history of the Milky Way and the Local Group, and into a coherent picture of compact objects and their environments, requires wide field, highly multiplexed spectroscopy that can reach faint stars with high information content (radial velocities, detailed abundances, multi-epoch coverage) over very large areas of sky. Existing and planned facilities either lack the necessary combination of aperture, multiplex, high-resolution capability and northern-hemisphere access to the key regions of interest, most notably the Galactic anti-centre and the bulk of the Local Group. GALATEA is a concept for a dedicated 15\,m-class, wide-field, multi-object spectroscopic facility in the northern hemisphere, with a corrected field-of-view of order $\sim 1~\mathrm{deg}^2$, $\sim 10{,}000$ fibres, and powerful medium- and high-resolution modes optimised for resolved stellar populations and products of stellar binary evolution. For the Milky Way, combining the astrometric baselines of \emph{Gaia} and \emph{GaiaNIR} with GALATEA’s deep, high-resolution spectroscopy will deliver a truly 6D, chemically resolved view of the outer disc, halo and disc--halo interface, as well as of the many classical and ultra-faint dwarf spheroidal satellites accessible from the northern hemisphere. For M31 and M33, GALATEA will play an analogous role by coupling its spectroscopy to the proper motions and deep imaging from \emph{Roman}~[8], in particular from programmes such as the proposed \emph{RomAndromeda} survey, enabling the first chemodynamical reconstruction of these spirals, their halos and their rich satellite systems on a par with the Milky Way. Existing northern multi-object spectroscopic facilities (APOGEE-2 [9], DESI [10], LAMOST [11], WEAVE [12]) operate on 2.5--4\,m-class telescopes and are therefore limited in depth and/or spectral resolution, while even Subaru/PFS~[13], the current benchmark on an 8--10\,m-class telescope, lacks the combination of survey depth, multiplex and dedicated observing time required for truly global chemodynamical mapping of the Milky Way and the Local Group. In contrast, GALATEA is conceived as a dedicated 15\,m, $\sim 1~\mathrm{deg}^2$, 10{,}000-fibre facility, providing an order-of-magnitude increase in effective survey power ($\propto$ $D^{2} N_{\mathrm{fibres}}$) over existing 4\,m-class wide-field facilities and substantial gains in collecting area and multiplex relative to PFS.

In the emerging landscape, GALATEA is therefore the natural northern wide field spectroscopic pillar, complementary to the planned Wide field Spectroscopic Telescope (WST)~[14] in the southern hemisphere and to the ultra-deep capabilities of 30--40\,m class ELTs. Together with WST, it would enable genuinely all sky, multi-epoch stellar spectroscopy, with GALATEA optimised for the Galactic anticentre and the Local Group, and WST for the southern Milky Way and cosmological surveys. At the same time, GALATEA would act as a survey feeder for ELTs — in particular TMT in the northern hemisphere — by delivering large, well-characterised samples of resolved stellar populations in the Milky Way, M31, M33 and their satellite systems, and by providing follow-up spectroscopy of compact-object candidates, interacting binaries and other products of stellar binary evolution, together with their stellar counterparts, identified by facilities such as \emph{LISA}~[6], \emph{NewAthena}~[7], \emph{Roman}/\emph{RomAndromeda}~[8] and other time-domain surveys. In this way, GALATEA helps to complete the chemodynamical and time-domain view of the nearby Universe initiated by these facilities.

\begin{figure}[!htbp]
\centering
\includegraphics[width=0.8\textwidth]{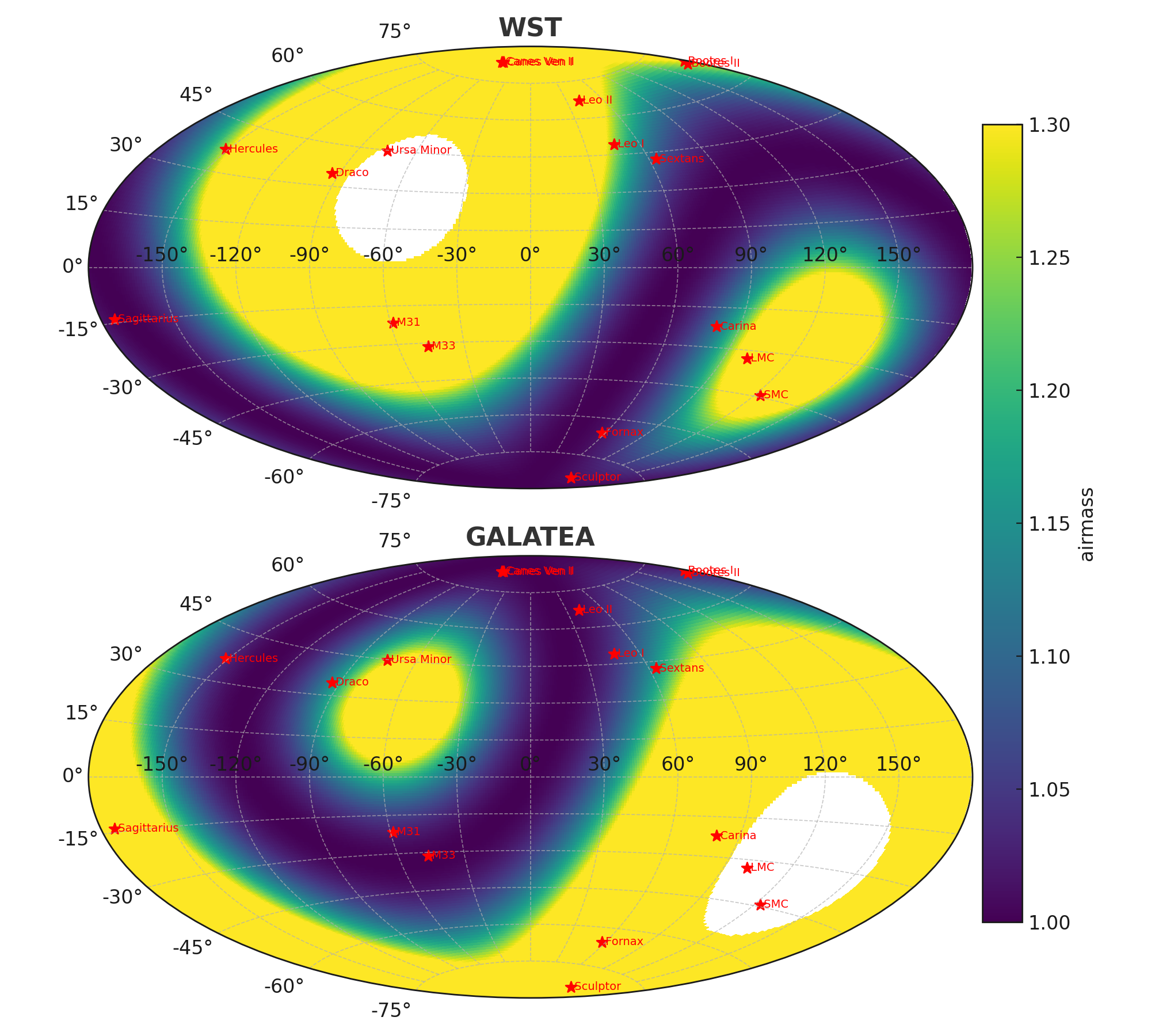}
\caption{\label{fig:wst_galatea_airmass}Comparison of the minimum airmass at transit for a WST-like southern site (top, $\varphi \simeq -25^\circ$) and GALATEA (bottom, $\varphi \simeq +30^\circ$), shown in Galactic coordinates (Aitoff projection, $\ell$ centred on the anticentre). Only the sky visible from each site is coloured; the colour scale indicates the minimum airmass in the range $1.0 \le X_{\min} \le 1.3$. Red stars mark the positions of M31, M33 and the known Milky Way dwarf spheroidal satellites, highlighting the complementary low-airmass access to Local Group targets from the two hemispheres.}
\end{figure}

\section{The Science Challenge}

\paragraph{The unsolved problem and the northern role.}
A central open problem in near field cosmology is to reconstruct the chemodynamical assembly of the Milky Way and the Local Group, from the warped, flared outer disc to the faintest halo substructures and dwarf satellites. These dynamical perturbations remain poorly understood: we still do not fully understand the origin of the outer-disc warp and flare in the Galactic anticentre, the nature of prominent halo substructures (e.g.\ Monoceros-like features, TriAnd like overdensities and associated streams), or the full extent and mass distribution of the haloes of M31 and M33 and their rich satellite systems. The key regions for addressing these questions --- the Galactic anticentre, the outer disc and halo in that direction, and most of the Local Group (M31, M33, their stellar streams and many of their dwarf companions) --- lie in the northern sky and cannot be mapped deeply and homogeneously from the south. Crucially, we still lack high-quality spectroscopy for the faint red-giant and main-sequence stars that carry most of the mass, angular momentum and merger history in these regions. The complementarity between North and South is illustrated in Fig.~\ref{fig:wst_galatea_airmass}, which compares the minimum airmass at transit for a WST like southern site and for GALATEA: from the South, low airmass access is naturally optimised for the classic southern dwarf spheroidals, whereas M31, M33 and most northern satellites are at high airmass or invisible. GALATEA, by contrast, provides low airmass coverage of the Galactic anticentre, M31, M33 and the bulk of the northern dwarf population, uniquely positioning it to deliver the first deep, homogeneous chemodynamical map of the northern half of the Local Group.

\paragraph{Scientific impact.}
Solving this problem will decisively advance our understanding of galaxy formation and hierarchical assembly. GALATEA will reveal how the Milky Way’s outer disc and halo were built, how warps, flares, thick discs and radial migration emerge, and how massive spirals like M31 and M33 assemble their haloes and satellite systems. A chemically resolved census of halo streams and Local Group dwarfs will deliver stringent tests of $\Lambda$CDM predictions for substructure and dark matter halo profiles, while large samples of extremely metal-poor stars in the outer Galaxy and in dwarf galaxies will pin down the earliest phases of chemical enrichment and the yields of the first supernovae. In parallel, combining GALATEA spectroscopy with Gaia (and GaiaNIR) astrometry and space-based asteroseismology will provide precise ages and chemodynamical characterisation for vast samples of stars, hitting the core goals of Galactic archaeology and linking exoplanet host stars directly to their birth environments, while statistically robust samples of binaries and compact remnants will open a new window on binary evolution, mass transfer, and the progenitors of stellar mass gravitational wave sources. Altogether, this programme cuts across cosmology and galaxy formation (the Local Group as a high-resolution benchmark), planetary science (exoplanets in their full Galactic context), and fundamental physics (dark matter, nucleosynthesis, asteroseismology, binary evolution, and the properties of the first stars).

\paragraph{Why this requires GALATEA.}
Answering these questions requires deep, high-resolution spectroscopy of faint stars over very large areas in the northern sky. In the northern hemisphere, the presence of M31 and M33 provides a strong scientific motivation for a larger aperture: increasing the aperture from 10\,m to 15\,m boosts the collecting area by a factor of 2.25, i.e.\ almost 0.9\,mag deeper at fixed exposure time and signal-to-noise. This is crucial to move beyond the very brightest red giants and reach fainter tracers in M31 and M33 (red clump and selected hot stars in their haloes), while simultaneously pushing towards the main-sequence turn-off in the outer halo of the Milky Way and in nearby dwarf galaxies. For a representative depth of $r \simeq 20.5$ towards the Galactic anticentre ($\ell \simeq 180^\circ$), GALATEA reaches main-sequence and turn-off stars with $M_r \simeq 4$--4.5 out to $d \simeq 16$--20 kpc ($R \simeq 24$--28 kpc), fully covering the warped and flared outer disc and the disc--halo interface. For more luminous tracers such as red clump and red giant stars ($M_r \simeq 0.5$ and $M_r \simeq -2$), the same limit corresponds to $R \sim 100$ kpc and $R \sim 300$ kpc, probing the outer halo. GALATEA’s 15\,m aperture, wide field-of-view and $\sim 10{,}000$ fibres enable contiguous chemodynamical mapping of key structures such as the anti-centre warp and flare, major halo streams, and the M31/M33 systems in a practical number of pointings, while the high-resolution spectrographs provide the detailed abundances and precise radial velocities needed for chemical tagging, orbit reconstruction and binary detection. These goals favour a wavelength coverage that includes both the near-UV/blue optical domain and a near-infrared extension. The near-UV/blue window contains many of the key diagnostic lines for chemical tagging and hot tracers, and is crucial for blue horizontal branch and blue straggler stars, which are intrinsically bright and, in the case of BHBs, excellent standard candles for dynamical mapping of the haloes of the Milky Way and M31/M33. At the same time, near-IR coverage close to the Galactic plane is essential to penetrate dusty, high extinction regions and to recover the chemodynamical structure of the warped, flared outer disc where optical surveys are severely limited. No existing or planned southern facility offers this combination of sensitivity, multiplex, and high-resolution optical+near-IR spectroscopy at faint limits for northern targets, making GALATEA uniquely suited to address this science challenge.


\section*{References}
{\footnotesize
\setlength{\parindent}{0pt}\noindent
[1] Prusti T. et al., 2016, A\&A, 595, A1,
[2] Hobbs D. et al., 2016, GaiaNIR White Paper, arXiv:1609.07325,
[3] Ivezi\'c {\v Z}. et al., 2019, ApJ, 873, 111,
[4] Laureijs R. et al., 2011, Euclid Study Rep., ESA/SRE(2011)12, arXiv:1110.3193,
[5] Rauer H. et al., 2014, Exp. Astron., 38, 249,
[6] Amaro-Seoane P. et al., 2017, LISA Consort., arXiv:1702.00786,
[7] Nandra K. et al., 2013, Athena+ White Paper, arXiv:1306.2307,
[8] Spergel D. et al., 2015, WFIRST/AFTA Rep., arXiv:1503.03757,
[9] Majewski S.~R. et al., 2017, AJ, 154, 94,
[10] Aghamousa A. et al., 2016, DESI Collab., arXiv:1611.00036,
[11] Zhao G. et al., 2012, Res. Astron. Astrophys., 12, 723,
[12] Dalton G. et al., 2016, Proc. SPIE, 9908, 99081G,
[13] Takada M. et al., 2014, PASJ, 66, R1,
[14] WST Consortium, 2024, WST Concept and Science Case, in prep.
\par}

\end{document}